\documentclass[aps,prl,twocolumn,amsmath]{revtex4-1} 
\pdfoutput=1
\usepackage{ulem}

\usepackage{color,graphicx}
\usepackage{bm}

\usepackage{amsmath}
\usepackage{amsfonts}
\usepackage{amssymb}

\usepackage{subfigure}
\usepackage[latin1]{inputenc}

\newcommand{\etal}{{\it et al.}}

\renewcommand{\-}{\,-\,}

\newcommand{\SRO}{Sr$_2$RuO$_4$ }
\newcommand{\SROdot}{Sr$_2$RuO$_4$}

\let\oldmarginpar\marginpar
\renewcommand\marginpar[1]{\-\oldmarginpar[\raggedleft\tiny #1]%
{\raggedright\tiny #1}}

\begin{document}


\title{Pairing symmetry and dominant band in \SROdot}
\author{Thomas Scaffidi}
\affiliation{Rudolf Peierls Centre for Theoretical Physics, Oxford OX1 3NP, United Kingdom}

\author{Jesper C. Romers}
\affiliation{Rudolf Peierls Centre for Theoretical Physics, Oxford OX1 3NP, United Kingdom}

\author{Steven H. Simon}
\affiliation{Rudolf Peierls Centre for Theoretical Physics, Oxford OX1 3NP, United Kingdom}

\date{\today}

\begin{abstract}
We study the superconductivity pairing symmetry in Sr$_2$RuO$_4$ in the limit of small interaction by extending a renormalization group calculation developed by Raghu {\it et al.} [Phys. Rev. B {\bf 81}, 224505 (2010)] to include spin-orbit coupling and multiband effects. 
We show these effects to be crucial to discriminate between the possible order parameters. 
In contrast to previous results and without the necessity of fine-tuning, we obtain pseudospin-triplet gaps of the same order of magnitude on the two-dimensional $\boldsymbol\gamma$ band and the quasi-one-dimensional $\boldsymbol\alpha$ and $\boldsymbol\beta$ bands.
The ratio of the gap amplitude on the different bands varies continuously with the interaction parameter.
The favored pairing symmetry is shown to be chiral when $\boldsymbol\gamma$ is slightly dominant and helical when $\boldsymbol\alpha$ and $\boldsymbol\beta$ are slightly dominant.
\end{abstract}

\maketitle

Strontium ruthenate \cite{maeno1994superconductivity,RevModPhys.75.657,JPSJ.81.011009} is a layered perovskite material exhibiting a transition at 1.5 K from a well-behaved Fermi liquid to a superconducting phase. 
Strong experimental evidence points towards an odd-parity order parameter (OP) \cite{ishida1998spin,PhysRevLett.85.5412,Nelson12112004,Kidwingira24112006}.
Based on multiple experiments\cite{luke1998time,PhysRevLett.84.6094,PhysRevLett.97.167002,Kidwingira24112006,1367-2630-12-7-075001,anwar2013anomalous}, the prevailing candidate for the symmetry of the OP has been the chiral $p$-wave state, ${\bf d} = (p_x \pm i p_y) \hat{{\bf z}}$, which breaks time-reversal symmetry (TRS), hosts topologically protected chiral edge states and is analogous to superfluid $^3$He-$A$ \cite{PhysRevLett.30.1108,PhysRev.123.1911}  (${\bf d}$ is defined below).

On the other hand, this state is supposed to carry edge currents at sample edges and domain walls, which have been elusive so far despite intense scrutiny \cite{PhysRevB.76.014526,PhysRevB.81.214501}.
As a result, other OP symmetries have been considered theoretically \cite{0953-8984-7-47-002,PhysRevB.73.134501,0295-5075-98-2-27010,JPSJ.82.063706}, including the helical states, ${\bf d} = p_x  \hat{{\bf x}} \pm p_y \hat{{\bf y}} $ and ${\bf d} = p_y  \hat{{\bf x}} \pm p_x \hat{{\bf y}} $.
These phases can be viewed as time-reversal invariant versions of chiral superconductors. 
Their edges host two counter-propagating Majorana modes of opposite spin whose net charge current is zero.

\begin{figure}[t!]
 \includegraphics[scale=0.25]{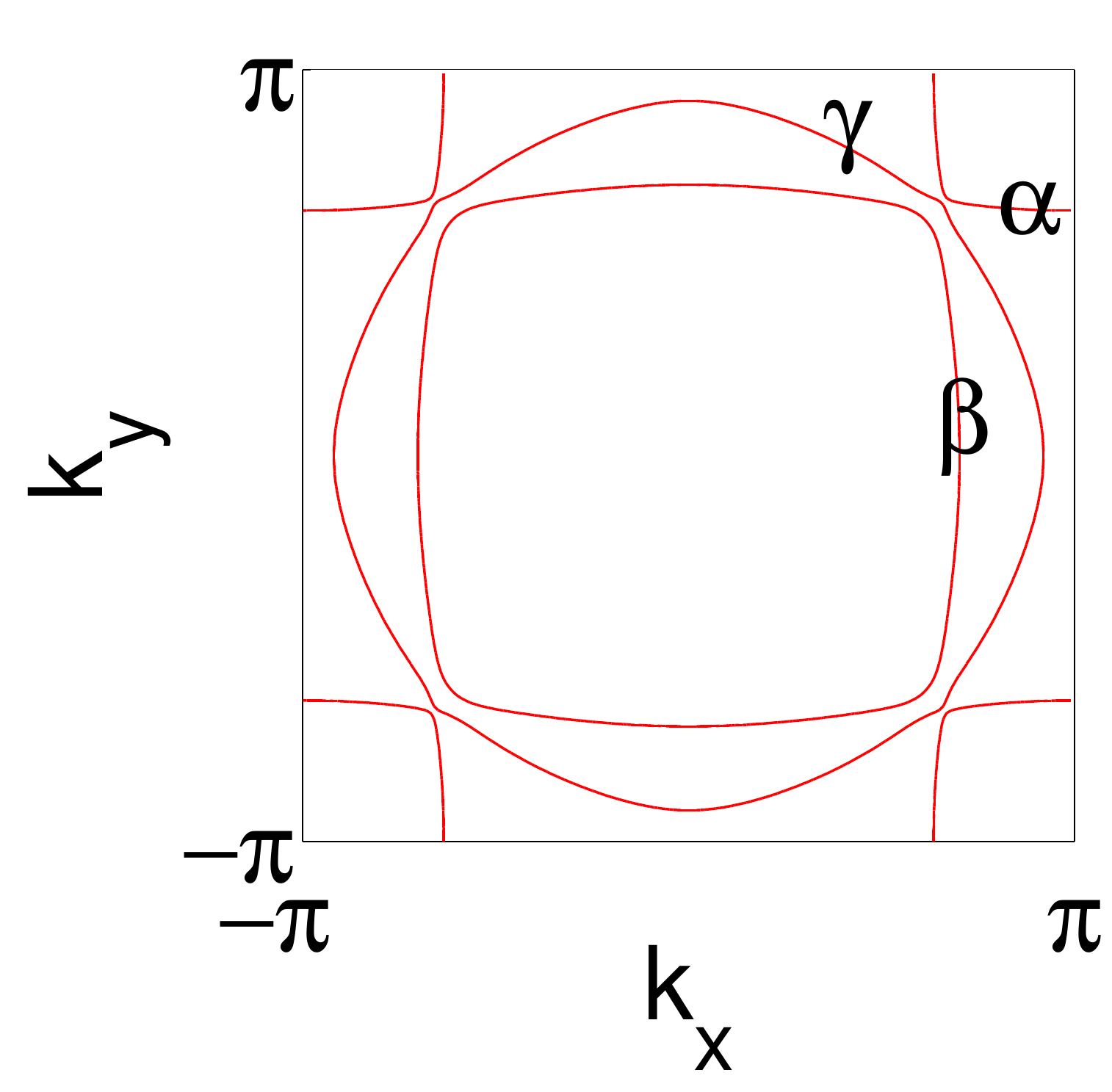}
\caption{Fermi surfaces for the tight-binding model $H$ given in Eq.~\eqref{hopping}.\label{bands}} 
\end{figure}

Another controversy has arisen recently regarding the band(s) on which the superconducting instability is dominant.
The Fermi surface (FS) of \SRO is made of three cylindrical sheets: The $\boldsymbol\gamma$ band is mainly derived from the Ru $4d_{xy}$ orbital and is fairly isotropic in the basal plane, while the $\boldsymbol\alpha$ and $\boldsymbol\beta$ bands are mainly derived from the Ru $4d_{xz}$ and $4d_{yz}$ orbitals and are quasi-one-dimensional (see Fig.~\ref{bands}). 

The prevailing assumption in the field has been that $\boldsymbol\gamma$ is the active band, due to its proximity to a Van Hove singularity.
This assumption was based on specific heat data \cite{PhysRevLett.92.047002}  and backed by several calculations \cite{JPSJ.69.3678,JPSJ.71.404,JPSJ.72.673,JPSJ.74.1818,0295-5075-104-1-17013} that predicted a dominant gap on $\boldsymbol\gamma$ and a subdominant, near-nodal gap on $\boldsymbol\alpha$ and $\boldsymbol\beta$.

This scenario was challenged recently. First, Raghu \etal \cite{PhysRevLett.105.136401}(see also \cite{PhysRevB.86.064525}) showed that, in absence of band coupling and in the weak-coupling limit, $\boldsymbol\alpha$ and $\boldsymbol\beta$ are the active bands. Second, Firmo \etal \cite{PhysRevB.88.134521} reported a phenomenological model with a gap amplitude of similar size on the three bands but slightly larger on $\boldsymbol\alpha$ and $\boldsymbol\beta$ than on $\boldsymbol\gamma$ that is consistent with specific heat and scanning tunneling microscopy (STM) measurements.

In this Rapid Communication, we extend the renormalization group (RG) scheme of Raghu \etal \cite{PhysRevB.81.224505,PhysRevB.83.094518,PhysRevB.85.024516,PhysRevB.88.064505,1742-6596-449-1-012031} by including spin-orbit coupling\cite{PhysRevLett.101.026406,PhysRevLett.112.127002} and multiband effects. 
This enables us to study the orientation of ${\bf d}$ at a microscopic level and determine the gap on the three bands. 
We find similarly sized gaps on the three bands without the necessity of fine-tuning.
Depending on the interaction parameter, we find two OPs that are compatible with the thermodynamic data: either a chiral gap whose amplitude is slightly larger on $\boldsymbol\gamma$, or a helical gap whose amplitude is slightly larger on $\boldsymbol\alpha$ and $\boldsymbol\beta$.


The three bands of strontium ruthenate are reproduced using the following tight-binding Hamiltonian for electrons hopping on a square lattice \cite{PhysRevLett.100.096601,0295-5075-49-4-473}
\begin{equation}
H = \sum_{{\bf k},s} \psi_s^{\dagger}({\bf k}) \hat{H_{s}}({\bf k}) \psi_s({\bf k}) 
\label{hopping}
\end{equation}
where $\psi_s({\bf k}) = [c_{{\bf k},A,s} ; c_{{\bf k},B,s} ; c_{{\bf k},C,-s}]^T$ with $s = 1$ ($-1$) for up (down) spins. The matrix $\hat{H}_s({\bf k})$ is given by \footnote{See Supplemental Material for more details about the form of the spin-orbit coupling in the Ru atomic orbitals basis.} 
\begin{equation}
\hat{H}_s({\bf k}) =
\begin{pmatrix}
E_{A} ({\bf k})  & g({\bf k}) - s i \eta &  i \eta \\
g({\bf k}) + s i \eta & E_{B} ({\bf k}) & - s \eta \\
 - i \eta & - s \eta & E_{C} ({\bf k}) 
\end{pmatrix} 
\end{equation}
where $E_{A} ({\bf k}) = -2 t \cos(k_x) - 2 t^{\perp} \cos(k_y) - \mu$, $E_{B} ({\bf k}) = -2 t^{\perp} \cos(k_x) - 2 t \cos(k_y) - \mu$, $E_{C} ({\bf k}) = -2 t' (\cos(k_x)+\cos(k_y)) - 4 t'' \cos(k_x) \cos(k_y) - \mu_c$ and $g({\bf k}) = -4t''' \sin(k_x) \sin(k_y)$.
$A$, $B$ and $C$ stand for the Ru orbitals $4d_{xz}$, $4d_{yz}$ and $4d_{xy}$ on each lattice site. The spin-orbit coupling (SOC) parameter is $\eta$ and the interorbital hopping term is $g({\bf k})$\footnote{Since both these parameters create repulsion between the bands, there is some freedom in their choice. 
Accordingly, our value of $t'''$ is smaller than in calculations without SOC \cite{JPSJ.71.404,0295-5075-104-1-17013} but is in agreement with a recent fit to ARPES data that includes SOC \cite{Zabolotnyy201348}.}. The parameters were chosen to reproduce the shape of the Fermi surfaces and the ratio of the effective masses of the different bands obtained from experiments\cite{RevModPhys.75.657,PhysRevLett.76.3786}: In dimensionless units, $(t,t^{\perp},t',t'',\mu,\mu_c,t''',\eta) = (1.0, 0.1, 0.8, 0.3, 1.0, 1.1, 0.01, 0.1)$.

After diagonalization, we obtain three pairs of degenerate pseudospin bands: 
\begin{equation}
H = \sum_{{\bf k},\alpha,\sigma} \epsilon_{{\bf k}, \alpha}  c_{{\bf k},\alpha,\sigma}^{\dagger}  c_{{\bf k},\alpha,\sigma} 
\end{equation}
 with  $\sigma = 1$ ($-1$) for $+$ ($-$) pseudospin and $\alpha = \boldsymbol\alpha, \boldsymbol\beta,\boldsymbol\gamma$. 
Roman indices refer to spin and orbital space while greek indices refer to pseudospin and band space.


We study the Coulomb interaction in the on-site $d$ atomic orbitals basis:
\begin{equation}
\begin{aligned}
H_{\text{int}} & = \sum_{i,a,s \neq s'} \frac{U}{2} n_{i a s} n_{i a s'} + \sum_{i,a\neq b,s,s'} \frac{U'}{2} n_{i a s} n_{i b s'} \\  & + \sum_{i, a\neq b,s,s'} \frac{J}{2} c^{\dagger}_{i a s} c^{\dagger}_{i b s'} c_{i a s'} c_{i b s} \\ &+  \sum_{i, a\neq b,s \neq s'} \frac{J'}{2} c^{\dagger}_{i a s} c^{\dagger}_{i a s'} c_{i b s'} c_{i b s}
\label{Hint}   
\end{aligned}
\end{equation}
where $i$ is the site index, $a=A,B,C$ is the orbital index, $\overline{s} \equiv -s$, $n_{i a s}\equiv c^{\dagger}_{i a s} c_{i a s}$, $U' = U - 2J $, and $J'=J$ \cite{Dagotto20011}.

%

Following Raghu \etal \cite{PhysRevB.81.224505}, we treat the weak-coupling limit, which corresponds to $U, J \ll W$ where $W$ is the bandwidth and $J/U$ a finite constant that fully parametrizes the interaction.
This is a well-controlled approximation in the sense that the solutions obtained are asymptotically exact in the weak-coupling limit. However, all real systems have finite interaction strengths and one is therefore forced to extrapolate this technique's results out of its strict regime of validity in order to make a link with experiments. Although this extrapolation probably leads to quantitative changes in our results, it should leave the qualitative trends untouched.

\begin{figure}[t!]
\includegraphics[width=0.5\columnwidth]{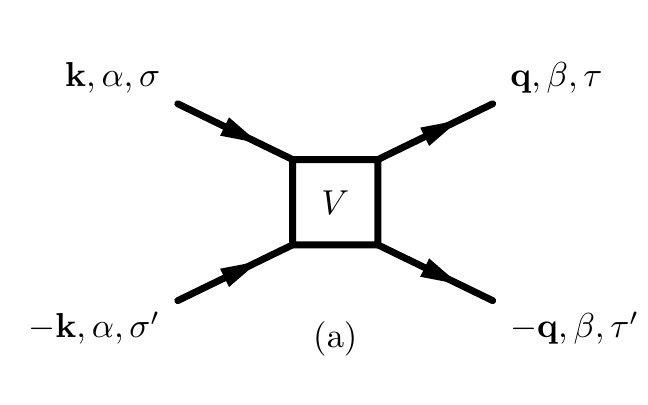}
 \includegraphics[scale=0.35]{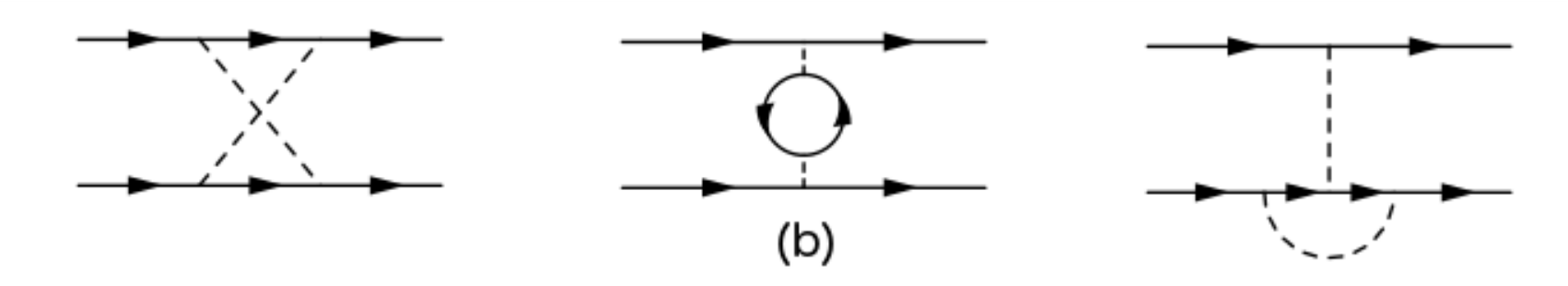}
\caption{(a) Diagram corresponding to the effective interaction $V({\bf k}_{\alpha}, {\bf q}_{\beta})$. (b) Nontrivial contribution to $V({\bf k}_{\alpha}, {\bf q}_{\beta})$ at one-loop order.\label{Diagrams}} 
\end{figure}

We integrate out all the modes with energies greater than an artificial cutoff to derive the effective particle-particle interaction in the Cooper channel $V({\bf k}_{\alpha}, {\bf q}_{\beta})$, where $\epsilon_{\alpha}({\bf k}_{\alpha})$ lies below the cutoff. 
The effective interaction $V({\bf k}_{\alpha}, {\bf q}_{\beta})$ corresponds to the diagram depicted in Fig.~\ref{Diagrams}(a). Its pseudospin dependence is left implicit for now.
Besides the bare vertex and its ladder, which give a trivial repulsive contribution, the effective interaction at one-loop order is made of the three diagrams shown in Fig.~\ref{Diagrams}(b). 
These diagrams are expressed in terms of the static susceptibility of the noninteracting system and correspond to the celebrated ``Kohn-Luttinger'' physics \cite{PhysRevLett.15.524,:/content/aip/proceeding/aipcp/10.1063/1.4818400}. 
The different bare vertices given in Eq.~\eqref{Hint} are represented diagrammatically by a unique dashed line that corresponds to a matrix in spin and orbital space. 
As the external propagators are in pseudo-spin and band space, the diagram expressions are supplemented by form factors from the unitary transformation going from spin and orbital to pseudospin and band space.

The second stage of the weak-coupling analysis is the calculation of the RG flow \cite{PhysRevB.81.224505}.
Each eigenmode of the effective interaction flows independently under the evolution of the running cutoff. 
These eigenmodes are solutions of
\begin{equation}
  \sum_{\beta}  \int_{\text{FS}}  \frac{d{\bf q}_{\beta}}{S_F} g({\bf k}_{\alpha},{\bf q}_{\beta}) \psi({\bf q}_{\beta}) = \lambda  \psi({\bf k}_{\alpha}) 
  \label{EigenMode}
\end{equation}
where
\begin{equation}
g({\bf k}_{\alpha},{\bf q}_{\beta})  = \sqrt{ \rho_{\alpha} \frac{\overline{v_{F,\alpha}}}{v_F({\bf k}_{\alpha})}} V({\bf k}_{\alpha}, {\bf q}_{\beta})  \sqrt{ \rho_{\beta} \frac{\overline{v_{F,\beta}}}{v_F({\bf q}_{\beta})}} \text{,}
\label{gDef}
\end{equation}
$S_F$ is the ``area'' of the FS, $\rho_{\alpha}$ is the density of states (DOS) of the band $\alpha$ at the Fermi level, and the average of the norm of the Fermi velocity is given by
\begin{equation}
\overline{v_{F,\alpha}}^{-1} =  \int  \frac{d{\bf k}_{\alpha}}{S_F} v_F({\bf k}_{\alpha}) ^{-1} \text{.}
\end{equation}
Since ${\bf k}_{\alpha}$ and ${\bf q}_{\beta}$ are constrained to lie on their respective FS, Eq.~(\ref{EigenMode}) is solved in matrix form once the FSs are discretized. 

The energy scale at which the perturbative treatment of the interaction breaks down corresponds to the critical temperature and is given by \cite{PhysRevB.81.224505}
\begin{equation}
T_c \sim W \exp\left(-\frac1{|\lambda|}\right)\text{.}
\end{equation}
The gap is proportional to the eigenvector \cite{PhysRevB.81.224505}:
\begin{equation}
 \Delta({\bf k}_{\alpha}) \sim  \sqrt{ \frac{v_F({\bf k}_{\alpha})}{\overline{v_{F,\alpha}} \rho_{\alpha}} } \psi({\bf k}_{\alpha}) \text{.}
\end{equation}

The pseudospin dependence of the order parameter is written in matrix form:
\begin{equation}
\begin{aligned}
 \Delta({\bf k}_{\alpha}) = 
\begin{pmatrix}
\Delta_{++}  & \Delta_{+-}    \\
\Delta_{-+} &\Delta_{--}   \\
\end{pmatrix} 
=
\begin{pmatrix}
-d_x+i d_y  & d_z +  \Delta_s   \\
d_z  - \Delta_s  &d_x+i d_y   \\
\end{pmatrix} 
\text{,}
\end{aligned}
\end{equation}
which defines a scalar order parameter $\Delta_s$ for the singlet case and a vectorial order parameter ${\bf d}$ for the triplet case. Since they are respectively even and odd under inversion, these two cases are mutually exclusive. The direction of ${\bf d}$ defines the normal to the plane in which the electrons are equal pseudospin paired.

The order parameter has to be in a given irreducible representation of the crystal symmetry group $D_{4h}$. The odd-parity representations can be split into two groups: the chiral state ${\bf d} = (p_x \pm i p_y) \hat{{\bf z}}$ and the helical states ${\bf d} = p_x  \hat{{\bf x}} \pm p_y \hat{{\bf y}} $ and ${\bf d} = p_y  \hat{{\bf x}} \pm p_x \hat{{\bf y}} $. The symbols $p_{x,y}$ stand for any function of momentum that has the same properties as $\sin(k_{x,y})$ under the symmetry operations of $D_{4h}$. The unit vectors $\hat{{\bf x}}$, $\hat{{\bf y}}$, and $\hat{{\bf z}}$ are the directions $a$ [100], $b$ [010], and $c$ [001].
The representation with the most negative pairing eigenvalue $\lambda$ corresponds to the favored state.

Since there is no consensus regarding the value of the interaction parameters\footnote{See Supplemental Material for a survey of estimates for these parameters found in the literature\cite{PhysRevB.86.165105, PhysRevLett.106.096401,PhysRevLett.110.167003,PhysRevB.86.045130}}, we will study {\it a priori} the whole acceptable range of $J/U$ and then compare predictions with experiments to infer its possible value.
The singlet case appears only for $J/U>0.29$ and can be discarded based on multiple measurements \cite{ishida1998spin,PhysRevLett.85.5412,Nelson12112004,Kidwingira24112006}.
While, for $J/U<0.065$, the chiral state is favored in agreement with the most prevailing assumption in the field, the helical state ${\bf d} = p_x  \hat{{\bf x}} + p_y \hat{{\bf y}} $ takes over for $0.065 < J/U < 0.29$. The helical state is the two-dimensional (2D) equivalent of superfluid $^3$He-$B$ \cite{PhysRev.131.1553}.

The TRS obeyed by the helical state is in contradiction with muon spin relaxation \cite{luke1998time} and optical Kerr effect \cite{PhysRevLett.97.167002} experiments but the interpretation of these experiments appears to conflict with the absence of edge currents \cite{0953-8984-21-16-164210,0034-4885-75-4-042501}. 
The absence of spin susceptibility decrease below $T_c$ for both in-plane and out-of-plane fields measured by NMR Knight shift experiments \cite{ishida1998spin,PhysRevLett.93.167004} has been interpreted as evidence in favor of a weakly pinned ${\bf d} \parallel c$ that can be rotated to the plane by a field $h \parallel c$ smaller than 20 mT.
We emphasize that a helical state with a weakly pinned ${\bf d} \perp c$ that would be rotated by a field $h \parallel ab$ smaller than 150 mT would also be consistent with these experiments.

Furthermore, the helical state would provide a simple explanation for the presence of edge states\cite{PhysRevLett.107.077003} but the absence of edge currents \cite{PhysRevB.76.014526,PhysRevB.81.214501}. 
It would also explain the emergence of out-of-plane spin fluctuations in the superconducting state \cite{PhysRevB.65.132507,doi:10.1143/JPSJ.78.074701}, which require in-plane fluctuations of ${\bf d}$. 
The disappearance of these fluctuations under an in-plane magnetic field would also be consistent with the expulsion of ${\bf d}$ from the plane under such a field.
Half-quantum vortices, measured recently in a mesoscopic sample of \SRO \cite{Jang14012011}, correspond to a spatially dependent rotation of ${\bf d}$ in order to accommodate a half-integer flux. They require a freeing of ${\bf d}$ from its intrinsic direction imposed by SOC and their existence is therefore equally plausible in the chiral and the helical state.
Given these contradictory experimental results, we will study these two states on an equal footing.

\begin{figure}[t!]
\begin{center}
\includegraphics[width=0.95\columnwidth]{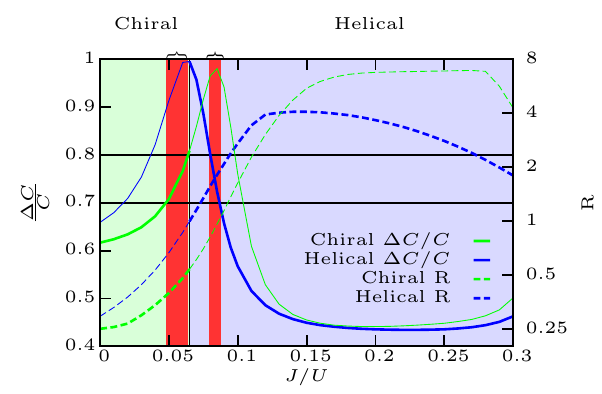}
\caption{
(Color online) Critical specific heat jump $\frac{\Delta C}{C}$ and ratio of the maxima of the gap amplitudes over the different bands $R=\frac{\max|\Delta_{\alpha,\beta}|}{\max|\Delta_{\gamma}|}$. The vertical line separates the stability regions of the chiral and helical OPs. The curve for a given OP is drawn in full width only in the OP's stability region. The horizontal lines delimit the range of $\frac{\Delta C}{C}$ estimated from experiments: $\frac{\Delta C}{C} = 0.75 \pm 0.05$ \cite{JPSJ.69.572,PhysRevB.88.134521}. The braces indicate the range of $J/U$ for which the prediction is in agreement with experiments.\label{SpecificHeat} 
}

\end{center}
\end{figure}

Once the mode with the most negative eigenvalue is identified, its eigenvector provides valuable information regarding the gap.
The gap scale is too small to be measured directly by angle-resolved photoemission spectroscopy (ARPES) but specific heat measurements have revealed properties of the order parameter \cite{PhysRevB.88.134521, JPSJ.73.1313}.
In Fig.~\ref{SpecificHeat}, we compare the measured \cite{JPSJ.69.572} critical jump in specific heat $\frac{\Delta C}{C}$ with its value calculated using BCS theory on the gap functions obtained from the RG technique.
The two highlighted regions correspond to a prediction for $\frac{\Delta C}{C}$ in agreement with experiments: the chiral OP at $J/U \simeq 0.06$ and the helical OP at $J/U \simeq 0.08$. 

The departure of $\frac{\Delta C}{C}$ from its well-known BCS maximal value of 1.43 measures the anisotropy of the gap over the three FS. 
A large difference between the scale of the gap amplitudes on the different bands corresponds to a value of $\frac{\Delta C}{C}$ that is smaller than experiments, as can be seen in Fig.~\ref{SpecificHeat}. Accordingly, the two predicted OPs in agreement with specific heat data have gaps of the same order on the three bands. The slightly dominant band is different in the two cases: The chiral state has a gap approximately two times larger on $\boldsymbol\gamma$ than on $\boldsymbol\alpha$ and $\boldsymbol\beta$, while the ratio of the helical gap amplitude on $\boldsymbol\gamma$ over the one on $\boldsymbol\alpha$ and $\boldsymbol\beta$ is approximately 0.7.
We checked that both these states give rise to a $T$ linear dependence of $C/T$ below $T_c$, in agreement with experiments \cite{JPSJ.69.572}.
By tuning $J/U$ towards smaller values, it is possible to obtain a largely dominant gap on $\boldsymbol\gamma$ like previously reported \cite{JPSJ.69.3678,JPSJ.71.404,JPSJ.72.673,JPSJ.74.1818,0295-5075-104-1-17013}.

\begin{figure}[t!]
\begin{center}
\includegraphics[width=0.95\columnwidth]{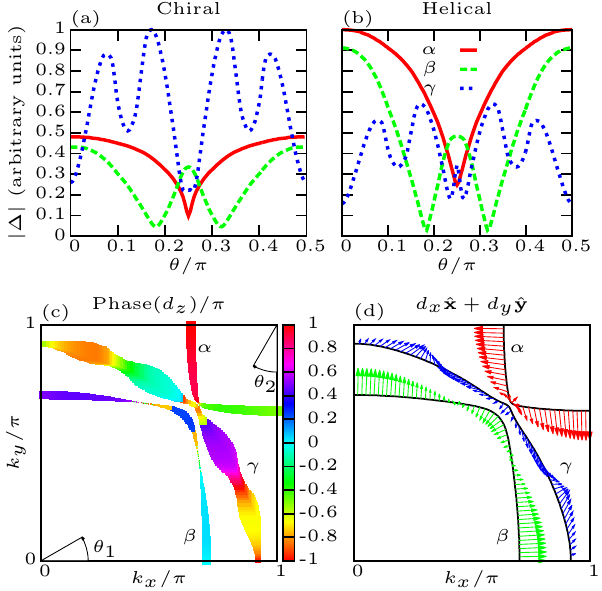}
\caption{(Color online) We represent the chiral OP ${\bf d} = d_z  \hat{{\bf z}}$  for $J/U=0.06$ (left) and the helical OP ${\bf d} = d_x  \hat{{\bf x}} + d_y \hat{{\bf y}}$ for $J/U=0.08$ (right). Panels (a) and (b) show the gap magnitude $|\Delta| \equiv \sqrt{\mathbf{d} \cdot \mathbf{d}^*}$. In panel (c), the color code gives the complex phase of $d_z$ around the three FSs. The width of the curve is proportional to $|\Delta|$. In panel (d), the vectors are proportional to $(d_x,d_y)$, where $d_x$ and $d_y$ are real. The angle $\theta$ refers to $\theta_1$ ($\theta_2$) in the case of $\beta$ and $\gamma$ ($\alpha$).\label{ThetaDelta}}
\end{center}
\end{figure}

As shown in Fig.~\ref{ThetaDelta}, the gaps on $\boldsymbol\alpha$ and $\boldsymbol\beta$ present near-nodes near the direction [110] in both cases. The incommensurate peak $\mathbf{Q}$ in the antiferromagnetic fluctuation spectrum \cite{PhysRevLett.83.3320} of these bands is known to be responsible for the appearance of these near-nodes \cite{PhysRevLett.105.136401,PhysRevB.88.134521}. 
As its fluctuations are mostly ferromagnetic, the $\boldsymbol\gamma$ band has been previously thought to host a fairly isotropic gap of the type $d_z = \sin(k_x) + i \sin(k_y)$, with only mild minima along [100] \cite{JPSJ.69.3678,JPSJ.71.404,JPSJ.72.673,JPSJ.74.1818,0295-5075-104-1-17013} and a complex phase increasing quasilinearly with $\theta$ (defined in Fig.~\ref{ThetaDelta}).
Interestingly, we find gap minima on $\boldsymbol\gamma$ along [110], which shows that the quasi-one-dimensional (quasi-1D) antiferromagnetic fluctuations peak $\mathbf{Q}$ is a source of anisotropy on this band as well.
Besides, the complex phase of our solution for $d_z$ in the chiral case [shown in Fig.~\ref{ThetaDelta}(c)] is a highly non-trivial and non-monotonic function of $\theta$.
Likewise, the in-plane orientation of ${\bf d}$ as a function of $\theta$ in the helical case [shown in Fig.~\ref{ThetaDelta}(d)] is much more involved than for the archetypal function ${\bf d} = \sin(k_x)  \hat{{\bf x}} + \sin(k_y) \hat{{\bf y}}$.

The rationale behind the association of the chiral state with a dominant $\boldsymbol\gamma$ and the helical state with dominant $\boldsymbol\alpha$ and $\boldsymbol\beta$ lies in the anisotropy of the normal-state spin dynamics. 
The chiral (helical) state has an out-of-plane (in-plane) $\mathbf{d}$ and is therefore driven by in-plane (out-of-plane) magnetic fluctuations. 
Due to SOC, the incommensurate peak $\mathbf{Q}$ is larger for the out-of-plane component of the susceptibility \cite{PhysRevB.65.220502}, thereby favoring a helical state when the quasi-1D bands are dominant. 
On the other hand, the (ferromagnetic) long wavelength part of the spectrum is larger for the in-plane component, which favors a chiral state when $\boldsymbol\gamma$ is dominant.

By a microscopic accounting of multiband and SOC effects, our model reconciles the two distinct scenarios of 2D superconductivity on $\boldsymbol\gamma$ versus quasi-1D superconductivity on $\boldsymbol\alpha$ and $\boldsymbol\beta$ inside one framework. 
As required by specific heat data\cite{PhysRevB.88.134521} and in contrast to previous RG calculations \cite{PhysRevLett.105.136401,0295-5075-104-1-17013}, similarly sized gaps on all three bands are obtained and, depending on the interaction parameter, the balance can be slightly tilted one way or another. 
As shown in Fig. \ref{SpecificHeat}, this result is true for both the chiral and the helical state and is therefore robust regardless of the favored pairing symmetry.

We now discuss experiments probing the relative size of the gaps on the different bands.
Recently, out-of-plane STM \cite{PhysRevB.88.134521} has exhibited the presence of a near-nodal gap of 0.350 meV on $\boldsymbol\alpha$ and $\boldsymbol\beta$. 
We find a position for the near-nodes on $\boldsymbol\alpha$ and $\boldsymbol\beta$ that is consistent with their phenomenological model, and we could reproduce their experimental tunneling DOS curves based on our gap.
Due to orbital anisotropy, the gap on $\boldsymbol\gamma$ cannot be measured with such an experiment.
The fact that the measured gap size corresponds to $2 \Delta / T_c \simeq 5$, which is close to the BCS value, was interpreted as evidence that $\boldsymbol\alpha$ and $\boldsymbol\beta$ are dominant. 
A gap 0.7 times smaller on $\boldsymbol\gamma$ was then inferred from the specific heat jump value, in agreement with our findings for the helical state.

On the other hand, the conductance of in-plane tunneling junctions \cite{PhysRevLett.107.077003} has been reported to present a two-step peak shape that is consistent with a dominant gap of 0.93 meV on $\boldsymbol\gamma$ and a subdominant gap of 0.28 meV on $\boldsymbol\alpha$ and $\boldsymbol\beta$. The relative sizes of the gap amplitude on the different bands would then point towards the chiral scenario.

The inclusion of $\eta$ is crucial to study the orientation of ${\bf d}$ since, without SOC, the spin $SU(2)$ symmetry would be preserved and the chiral and helical states would be degenerate. 
The splitting between the pairing eigenvalue of these states grows with the magnitude of $\eta$ but our conclusions are robust against a change in this parameter: The favored state is always chiral with a (slightly) dominant $\boldsymbol\gamma$ for small $J/U$ and helical with (slightly) dominant $\boldsymbol\alpha$ and $\boldsymbol\beta$ for larger $J/U$ (see Supplemental Material for more details).

Finally, we emphasize the need for new experiments that would make it possible to discriminate between the two proposed states. 
In-plane STM could be one of them since it could also measure the gap on $\boldsymbol\gamma$ unlike in the out-of-plane case. 
Experiments probing the phase of the order parameter, including quasiparticle interference and Josephson tunneling spectroscopy\cite{Kidwingira24112006,1367-2630-12-7-075001,anwar2013anomalous}, could be discriminating but their interpretation is nontrivial given the reported convoluted dependence of that phase on the in-plane orientation.
Methods to detect helical edge modes have also been proposed recently \cite{PhysRevB.85.140501}.

\begin{acknowledgments}
Helpful conversations with Sri Raghu, Steve Kivelson, Suk Bum Chung, Andy Mackenzie, Andrew Green, Jonathan Keeling, Chris Hooley, Clifford Hicks, Ed Yelland, Jim Sauls, Catherine Kallin, Andrea Damascelli and Peter Hirschfeld are acknowledged. This work is supported by EPSRC Grants No. EP/I032487/1 and No. EP/I031014/1, the Clarendon Fund Scholarship, and the University of Oxford.
\end{acknowledgments}

\bibliography{PRB_SM}

\section{Supplemental Material}
\subsection{Spin-orbit coupling}
The spin-orbit coupling acts as an on-site term, $H_{\text{SOC}} = 2 \eta \sum_i {\bf L}_i \cdot {\bf S}_i$, where the sum is over the Ru sites.
The crystal field splits the five Ru d orbitals in the $e_g$ doublet and the $t_{2g}$ triplet but only the $t_{2g}$ orbitals are relevant close to the Fermi level. These three orbitals behave like a $l=1$ angular momentum representation. 
Once expressed in terms of these orbitals only, the spin-orbit coupling Hamiltonian becomes \cite{PhysRevLett.100.096601,0295-5075-49-4-473}
$$ H_{\text{SOC}} = i \eta \sum_{\bf k} \sum_{l,m,n} \epsilon_{l m n} \sum_{s,s'} \sigma^{n}_{s s'} c^{\dagger}_{{\bf k} l s} c_{{\bf k} m s'} $$
where $l,m,n$ are orbital indices, $s,s'$ are spin indices and $\sigma^n$ is the $n$-th Pauli matrix. The orbital indices are defined in the following way: $l=1,2,3$ for, respectively, the orbital $d_{yz}$ (B), $d_{zx}$ (A) and $d_{xy}$ (C).

\subsection{Interaction parameters}
In Table \ref{JU}, we give different estimates of the interaction parameters used in the main text ($U$, $U'$ and $J$) that can be found in the literature.
In Refs. \cite{PhysRevB.86.165105} and \cite{PhysRevLett.106.096401}, a constrained random phase approximation (cRPA) calculation was performed to estimate these parameters. These two references give consistent results and an estimate for $J/U$ of 0.1. 
In Ref. \cite{PhysRevLett.110.167003}, an RG calculation performed in the one-dimensional limit of the $d_{zx}$ and $d_{yz}$ orbitals lead to the right prediction for the crossover to 3D Fermi liquid behaviour in \SRO($T_{3D}\simeq 60K$). A value of 2.2 eV was taken for $U$ and the relevant parameter range for $J$ was considered to be between $0.13$ and $0.4$ eV. This corresponds to a value of $J/U$ between $0.059$ and $0.18$. 
These estimates are in fair agreement with the range of $J/U$ for which our calculation is in agreement with the measured critical specific heat jump, which is roughly given by $0.05<J/U<0.065$ and $0.075<J/U<0.085$. 
In Ref. \cite{PhysRevB.86.045130} (see also references therein), a mean-field (MF) rotationally invariant slave bosons calculation was performed to study the impact of the Coulomb repulsion on the quasiparticle bands. From a survey of numerous references, they located $U$ in the region 1.5-3.1 eV and $J$ at 0.35 eV. 
Finally, Ref. \cite{PhysRevLett.106.096401} also reports a local density approximation associated with a dynamical mean field theory (LDA+DMFT) calculation. They obtain an estimate of $J=0.4$ eV by fitting the predicted mass enhancement to the experimental value. This estimate is somewhat larger than the previous ones.

\begin{table}[h]
\centering
    \begin{tabular}{|c| c |c |c |c|c|c|}
    \hline
    Ref.  & Method &$U$ & $U'$ & $J$ & $J/U$ \\ \hline
    \cite{PhysRevB.86.165105} & cRPA & 2.56 & 1.94 & 0.26 &  0.101\\
    \cite{PhysRevLett.106.096401} & cRPA &2.3 & $U-2J$ & 0.25 &  0.108\\    
    \cite{PhysRevLett.110.167003} &1D RG& 2.2 & - & 0.13-0.4 & 0.059-0.18 \\
    \cite{PhysRevB.86.045130} & MF & 1.5-3.1 & $U-2J$ & 0.35 & 0.11-0.23 \\
    \cite{PhysRevLett.106.096401} & LDA+DMFT & 2.3 & $U-2J$ & 0.4 &  0.17\\


    \hline

    \end{tabular}

    \caption{Interaction parameters (in eV) obtained by various methods. }
            \label{JU}
\end{table}

\subsection{Pairing eigenvalue}

\begin{figure}[h!]
\begin{center}
\includegraphics[width=0.95\columnwidth]{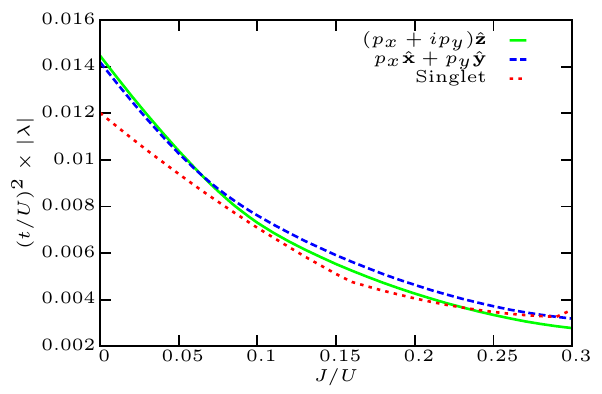}
\caption{
RG eigenvalue $\lambda$ for the chiral state, the most favoured helical state and the singlet state for the parameters given in the main text.\label{JLambda} 
}

\end{center}
\end{figure}

In Fig.~\ref{JLambda}, we show the pairing eigenvalue $\lambda$ for different pairing symmetries. The favoured state is the one with the largest value of $|\lambda|$. We show the eigenvalue for two odd-parity channels: one for the chiral state ${\bf d} = (p_x \pm i p_y) \hat{{\bf z}}$ and one for the most favoured helical state  ${\bf d} = p_x  \hat{{\bf x}} + p_y \hat{{\bf y}} $. We also show the most favoured state in the even-parity (i.e. pseudo-spin singlet) channel. Except for very high $J/U$ ($>0.29$), the even-parity channel is never favoured. 

\begin{figure}[h!]
\begin{center}
\includegraphics[width=0.95\columnwidth]{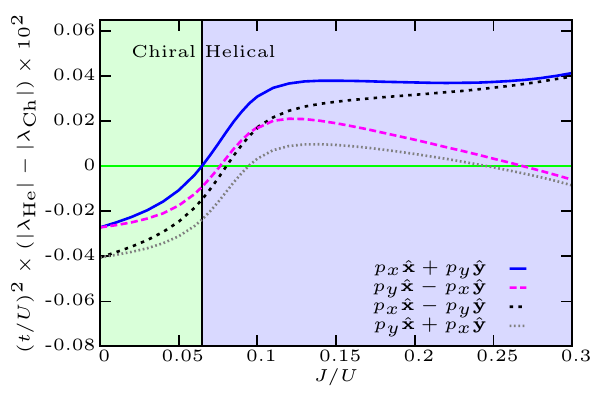}
\caption{
Splitting of the RG eigenvalue $\lambda$ between the four different helical states and the chiral state for the parameters given in the main text. \label{Splitting} 
}

\end{center}
\end{figure}

The splitting between the pairing eigenvalue of the helical states and the chiral state is shown in Fig.~\ref{Splitting}. The chiral state is favoured for $J/U < 0.065$ while the helical state  ${\bf d} = p_x  \hat{{\bf x}} + p_y \hat{{\bf y}} $ takes over for $J/U > 0.065$.

\subsection{Magnitude of spin-orbit coupling}
\begin{figure}[h!]
\begin{center}
\includegraphics[width=0.95\columnwidth]{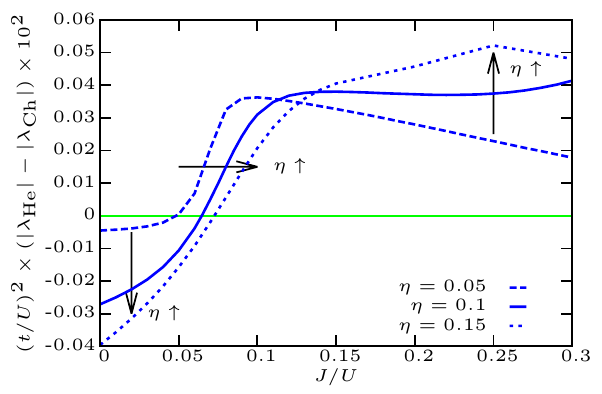}
\caption{
Splitting of the RG eigenvalues $\lambda$ between the chiral and the helical state for different SOC parameters $\eta$. All the other parameters are given in the main text.\label{SplittingLambda} 
}

\end{center}
\end{figure}

The impact of $\eta$ on the splitting between the pairing eigenvalue of the helical state and the chiral state is shown in Fig.~\ref{SplittingLambda}. 
The curves are fairly similar for all values of $\eta$: a region of negative splitting (i.e. favoured chiral state) at small $J/U$ and a region of positive splitting (i.e. favoured helical state) at large $J/U$. These two regions are linked by a cross-over at a certain value for $J/U$. As figured by the arrows in Fig.~\ref{SplittingLambda}, increasing $\eta$ does mostly two things: it increases the amplitude of the splitting (be it positive or negative) in the two aforementionned regions and slightly increases the value of $J/U$ at which the cross-over happens. It also makes the cross-over smoother. 
In the limit $\eta \rightarrow 0$, the splitting would go to zero, from below in the former region and from above in the latter. 

\begin{figure}[h!]
\begin{center}
\includegraphics[width=0.95\columnwidth]{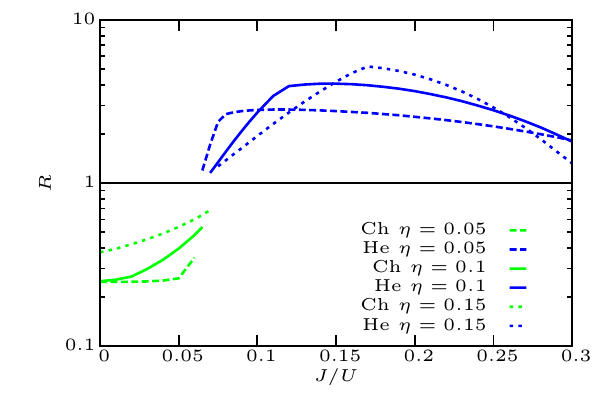}
\caption{
Ratio of the maxima of the gap amplitudes over the different bands $R=\frac{\max|\Delta_{\alpha,\beta}|}{\max|\Delta_{\gamma}|}$ for different SOC parameters $\eta$. All the other parameters are given in the main text. At each value of $J/U$, only the curve for the most favoured state (chiral or helical) is shown.``Ch'' stands for chiral and``He'' stands for helical. \label{RatioMaxima} 
}

\end{center}
\end{figure}

The ratio of the maxima of the gap amplitudes over the different bands $R=\frac{\max|\Delta_{\alpha,\beta}|}{\max|\Delta_{\gamma}|}$ for different SOC parameters $\eta$ is shown in Fig.~\ref{RatioMaxima}. Regardless of the value of $\eta$, the chiral state favoured at small $J/U$ has a larger gap magnitude on $\boldsymbol\gamma$ while the helical state at larger $J/U$ has a larger gap amplitude on $\boldsymbol\alpha$ and $\boldsymbol\beta$.

In summary, the amplitude of $\eta$ does not modify qualitatively our findings of a favoured chiral state with a (slightly) dominant $\boldsymbol\gamma$ for small $J/U$ and a favoured helical state with (slightly) dominant $\boldsymbol\alpha$ and $\boldsymbol\beta$ for larger $J/U$.

\end{document}